\documentclass[epjCONF,columns]{svjour}
\usepackage{graphics}
\usepackage[varg]{txfonts} 
\usepackage{subfigure}
\usepackage[latin1]{inputenc}
\session-title{Hadron Collider Physics Symposium 2011}

\newcommand{\ETmiss}{\ensuremath{E_{\rm T}^{\rm miss}}}
\begin{document}
\title{Search for Single Top tW Associated Production in the Dilepton Channel at CMS}
\author{Jochen Ott on behalf of the CMS Collaboration}
\institute{Karlsruhe Institute of Technology}
\abstract{
We present a first study of the single top quark W-associated production 
(tW) in proton-proton collisions at the LHC at a centre-of-mass
energy of 7TeV, using data collected with the CMS experiment.
The search is performed in the dileptonic final 
states ee/e$\mu$/$\mu\mu$ with a selection based on kinematical properties and b-tagging information.
The contribution of the Z+jets processes 
to the background is estimated from a sideband in data. Two $\rm t\bar t$ dominated control regions 
are used to constrain the normalization of top quark pair production in 
the signal region. An excess of events over the expected background is observed. Assigning
this excess to events from tW production, the extracted tW cross section is in agreement
with the Standard Model expectation.
} 
\maketitle

\section{Single Top tW Production}
\label{sec:prod}

\begin{figure}
\centering
\subfigure[]{
\resizebox{0.48\columnwidth}{!}{\includegraphics{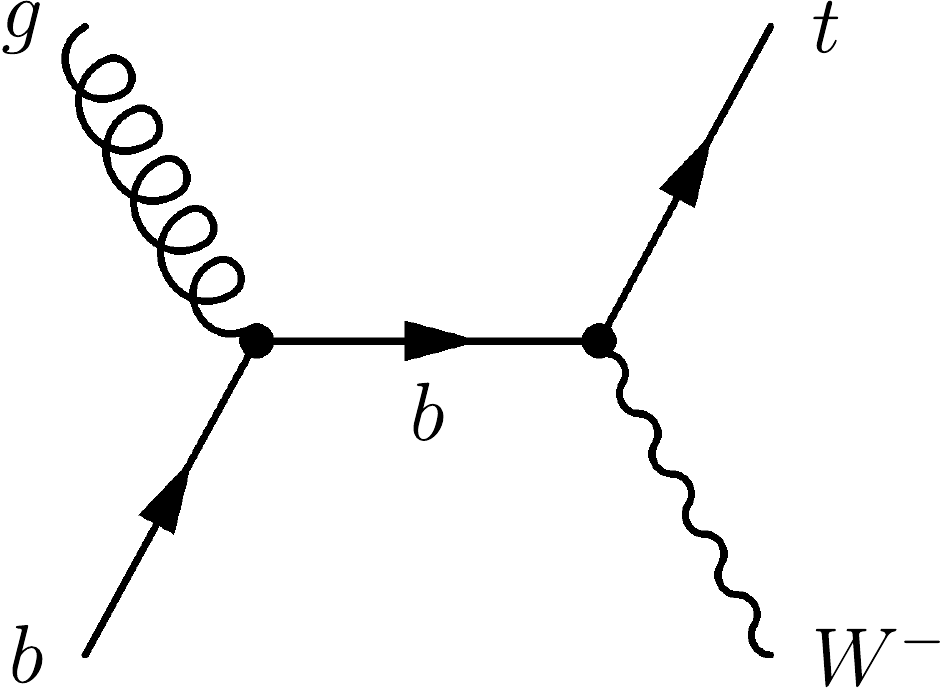}}}
\subfigure[]{
\resizebox{0.48\columnwidth}{!}{\includegraphics{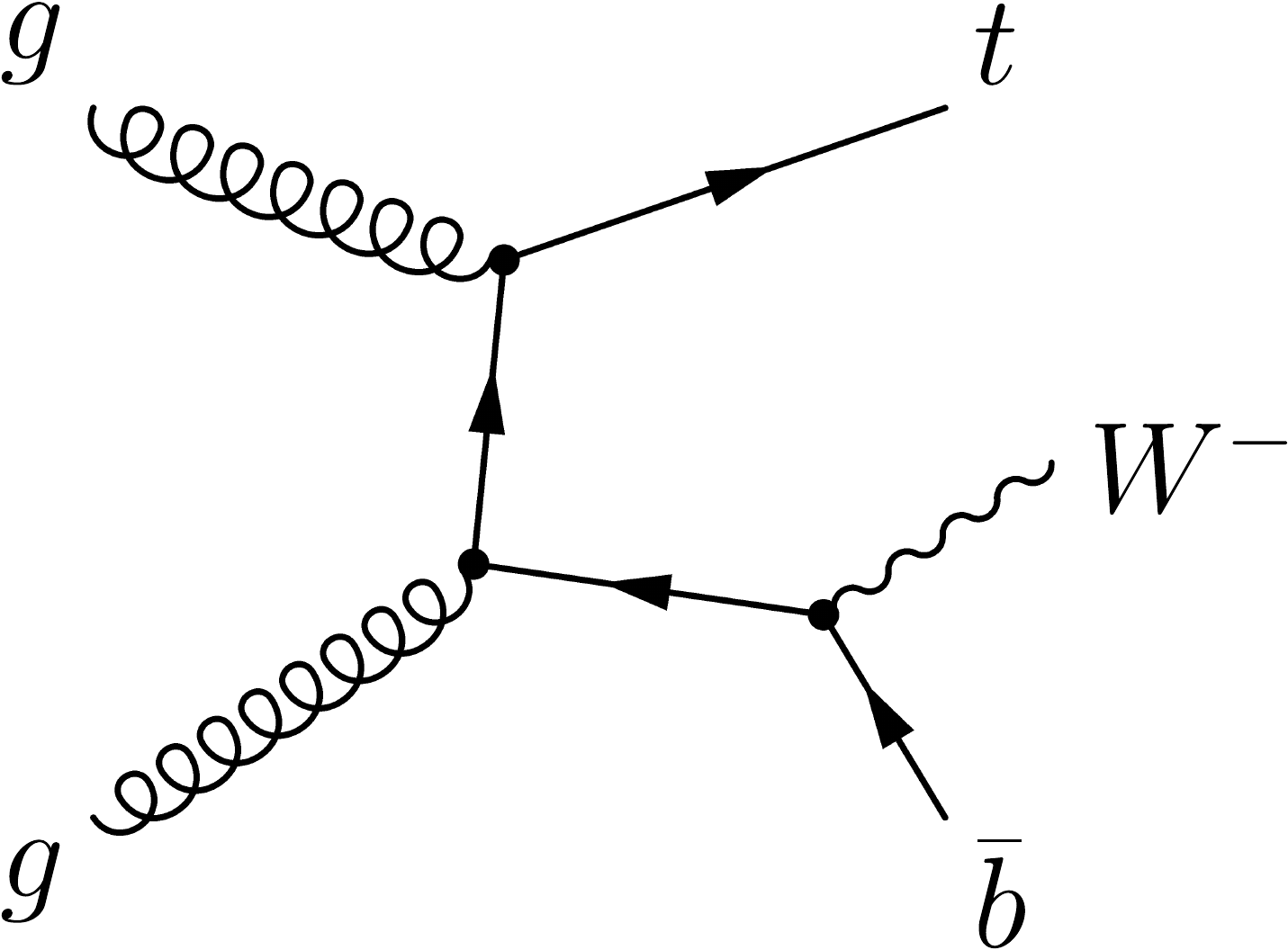}}
}
\caption{(a) Leading order Feynman diagram for tW production. (b)
Next-to-leading order Feynman diagram for tW production which also appears as a leading order diagram
for top quark pair production (and partial decay).}
\label{fig:feyn}
\end{figure}

Single top quark associated production (tW) is characterized by a final state with a top quark and a W boson.
A contributing leading-order Feynman diagram is shown in Fig.\ \ref{fig:feyn}~(a).
At next-to-leading order, real contributions with an additional bottom quark in the final state lead to
ambiguities in the conceptual distinction between the tW process and top quark pair production, as shown
in Fig.\ \ref{fig:feyn}~(b). Two possibilities to resolve this ambiguity~\cite{drds} have been considered: in the
\emph{diagram removal} method, all diagrams which can also arise in top quark pair production are removed for the calculation.
This is used as default scheme for the simulation used in this analysis.
The second possibility is the \emph{diagram subtraction} method
which locally subtracts resonant contributions. The difference between these two methods
is considered as a systematic uncertainty.

The predicted Standard Model cross section evaluated at approximate NNLO is $15.6 \pm 1.2$~pb~\cite{kidonakis:st}.

\section{Event Selection}
\label{sec:sel}
The present analysis uses a dataset of proton-proton collisions at $\sqrt{s}=7$~TeV which
corresponds to an integrated luminosity of 2.1~fb$^{-1}$, recorded with the CMS detector~\cite{cms}.

This analysis considers tW dilepton events in which the final state W boson and the W boson from
the top quark decay both decay into a charged lepton (e/$\mu$) and a neutrino. The event signature consists of
2 oppositely charged leptons, missing transverse energy from the neutrinos, and one b-jet from the top quark decay.

Events are selected online by a corresponding dilepton trigger (ee/e$\mu$/$\mu\mu$). In the offline
analysis, events are required to have exactly two isolated, oppositely charged leptons with $p_{\rm T} > 20$~GeV
and $|\eta| < 2.4$ (2.5) for muons (electrons). To reduce the number of selected Z+jet events in the ee and $\mu\mu$
channels, events with invariant dilepton mass $m_{\ell\ell}$ between 81 and 101 GeV are rejected. To further
reduce events with no prompt neutrinos (Z+jets, QCD), a
cut on missing transverse energy, $\ETmiss > 30$~GeV is placed in the ee and $\mu\mu$ channels.

As tW signal has a b-quark in its final state,
signal events are expected to have exactly one b-tagged jet while for the dominating background process,
top quark pair production, two b-tagged jets are expected. To simultaneously extract signal and constrain the
top quark pair background, three event categories are defined based on jet and b-tag multiplicity:
the ``1jet 1tag'' category contains events with exactly one jet with $p_{\rm T} > 30$~GeV and $|\eta| < 2.4$
which has been identified by a b-tagging algorithm based on the reconstruction of a secondary vertex~\cite{btagging}.
Similary, the ``2jet 1tag'' and ``2jet 2tag'' categories contain events with exactly 2 jets and where
either one or both of them are b-tagged.

\begin{figure}
\centering
\subfigure[]{
\resizebox{0.7\columnwidth}{!}{\includegraphics{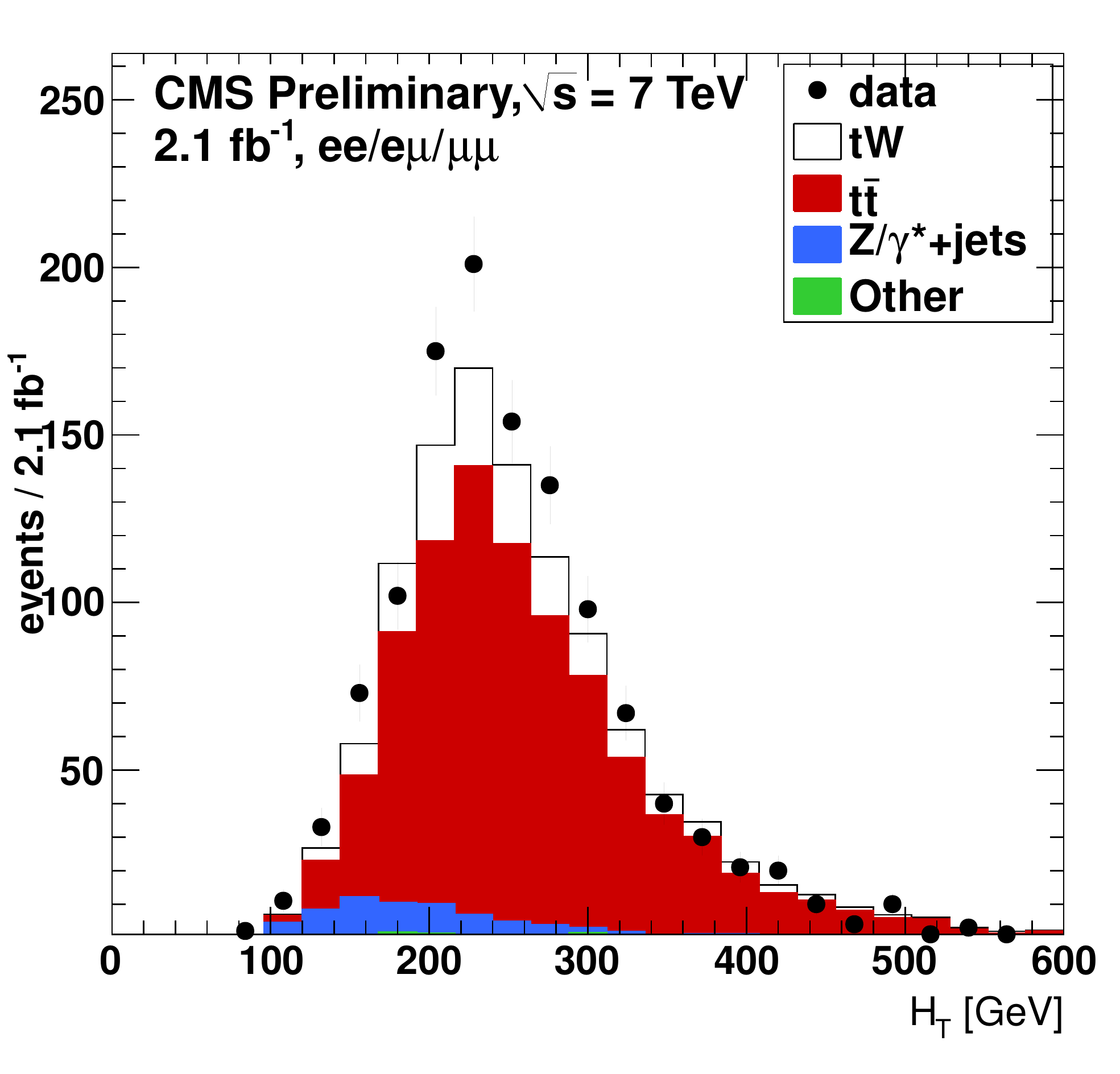}}
}
\subfigure[]{
\resizebox{0.7\columnwidth}{!}{\includegraphics{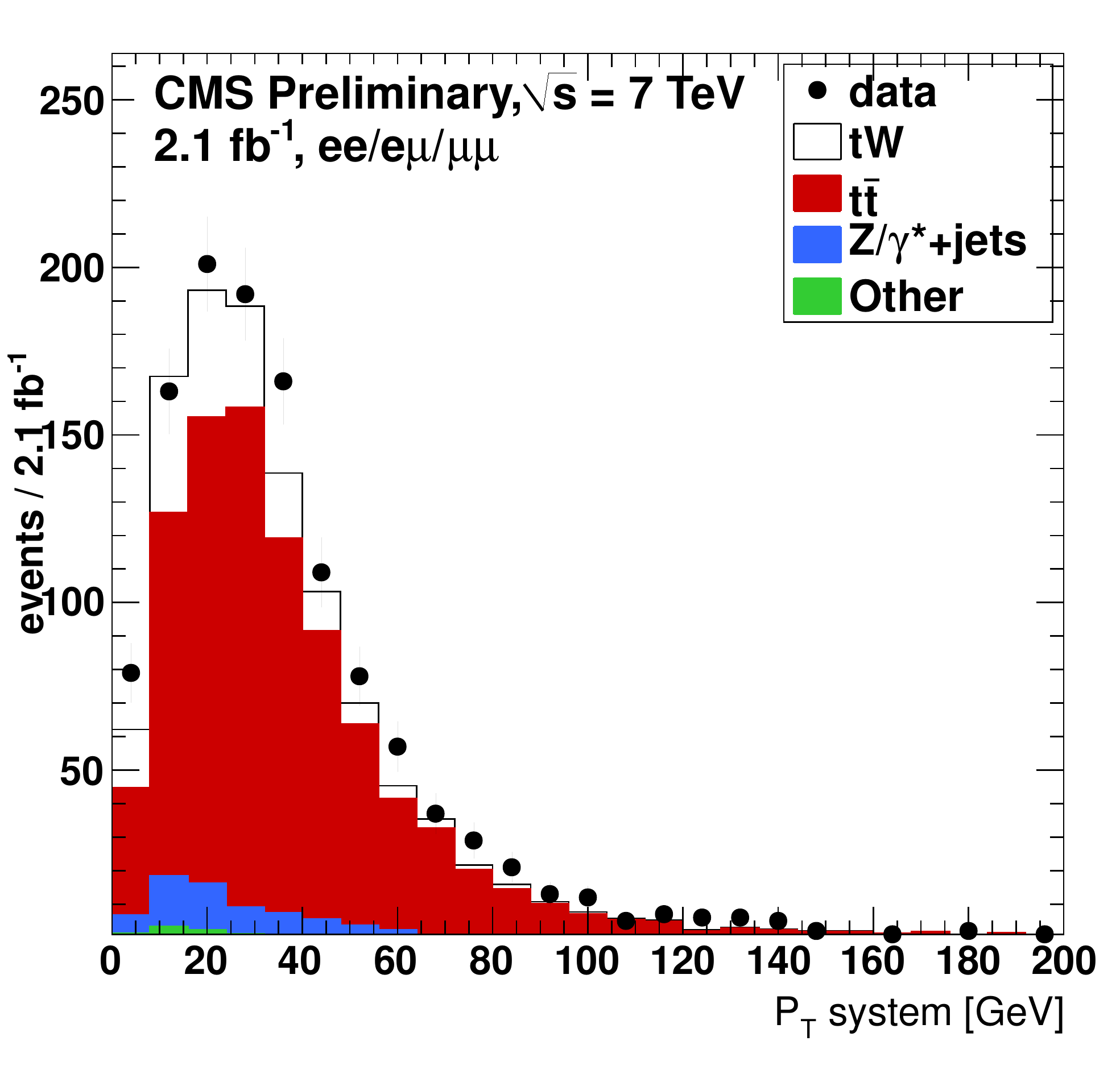}}
}
\caption{Variables used in the event selection for events with one b-tagged jet:
(a) $P_{\rm T}^{\rm system}$, defined as the transverse momentum sum of jets, leptons and 
missing transverse energy for events with exactly. (b) $H_{\rm T}$, defined as the scalar sum
of the leptons, jets and missing transverse energy.\label{fig:ptht}}
\end{figure}

Most of the tW signal is in the 1jet 1tag category. In this selection,
the dominating background are top quark pair events in the dilepton channel
in which one of the two expected jets is outside the kinematic acceptance. This leads to a momentum imbalance
in the variable $P_{\rm T}^{\rm system}$ which is defined as the the vector-sum of transverse momenta of the
leptons, missing transverse energy, and the jet: while for tW, this is the transverse momentum of the complete
final state which is expected to be small, it is expected to have larger values for $\rm t\bar t$
due to the missing jet. The $P_{\rm T}^{\rm system}$ variable is shown in Fig.\ \ref{fig:ptht}~(a). Events
with $P_{\rm T}^{\rm system} < 60$~GeV are selected for further analysis.

In the e$\mu$ channel, where no invariant mass and $\ETmiss$ requirements are applied, $H_{\rm T}$
(Fig.\ \ref{fig:ptht}~(b)), defined as the scalar sum of the transverse momenta of the leptons, jets,
and $\ETmiss$, is required to be larger than 160 GeV.

After this selection, the dominating background process is top quark pair production. In the same-flavor lepton
channels ee and $\mu\mu$, there is a non-negligible contribution from Drell-Yan events. Other background processes
include W+jets with a fake lepton and diboson processes (WW/WZ/ZZ).

\section{Background Estimation}
\label{sec:bkg}
The background contributions from W+jets and diboson processes are expected to be very small and taken
from simulation.

To estimate the background contribution from top quark pair production, simulated events are used, scaled to
the approximate NNLO cross section calculation from~\cite{kidonakis:ttbar}. The top quark pair cross
section is allowed to vary in the statistical evaluation within the systematic uncertainties.

To determine the remaining number of Drell-Yan background events, a technique based on a data sideband is used
which uses the number of events in the Z boson mass peak, $N^{\rm{obs}}_{\ell\ell,\rm{in}}$,
defined as the number of events in the $m_{\ell\ell}$ cut region from 81 to 101~GeV.
From this number, the non-resonant contribution
from processes with two final state W bosons (such as $\rm t\bar t$, tW, WW), is subtracted which
is estimated in the $e\mu$ channel. Finally, the predicted ratio of events inside and outside
the veto region is used to estimate the remaining number of Drell-Yan events passing the selection:
\[
N_{\rm{est}}^{\rm{DY}} = \frac{N_{\ell\ell,\rm{out}}^{\rm{MC}}}{N_{\ell\ell,\rm{in}}^{\rm{obs}}} \cdot
 (N_{\ell\ell,\rm{in}}^{\rm{obs}} - \frac 1 2 k N_{e\mu} )
\]
where $k$ corrects for differences in e/$\mu$ reconstruction efficiency and the factor $\frac 1 2$ accounts for
the ratio of branching ratios of same-flavor to different-flavor lepton channels
in dilepton tW and top quark pair events, BR($\ell\ell$) / BR(e$\mu$).

\section{Systematic Uncertainties}
\label{sec:syst}
A number of different sources of uncertainty affect the expected background and signal yield.
The considered uncertainties are
\begin{itemize}
\item \textbf{Pileup modeling}: additional proton-proton interactions in the same bunch crossing (pileup) shift 
 lepton isolation and jet energies. A possibly imperfect modeling of pileup
 introduces this uncertainty which turns out to be below 1\% for all yields.
\item \textbf{Trigger efficiency}: the efficiency of the online selection is known to 1.5\%.
\item \textbf{Lepton reconstruction and identification efficiency}: the efficiencies are known to 1\% (2\%) for
muons (electrons).
\item \textbf{\ETmiss modeling}: to account for the uncertainty of ca\-lo\-rimeter response of energy not included in jets and
leptons, this energy is smeared by 10\% which changes the signal acceptance by 1--2\%.
\item \textbf{Jet energy scale and resolution}: the absolute jet energy calibration is known to about 2--3\%~\cite{jes},
the jet energy resolution to about 10\%~\cite{jer}. Varying jet energies according to these uncertainties yields to acceptance
differences of about 1--2\% for tW signal and 4--6\% for $\rm t\bar t$ background.
\item \textbf{Background normalization}: the estimate for Drell-Yan background events is assigned an uncertainty of 50\%,
for other backgrounds, the cross section uncertainties from theory are used.
\item \textbf{tW and $\rm t\bar t$ modeling}: a number of parameters in the simulation, such as the factorization
and renormalization scale, matrix-element / parton shower matching parameters, parton distribution functions, diagram
subtraction and removal methods, have been varied in the simulation of tW signal and $\rm t\bar t$.
\item \textbf{Luminosity}: the calibration of the absolute integrated luminosity is 4.5\%.
\item \textbf{B-tagging}: the efficiency of the b-tagging algorithm to correctly tag a b-jet is known
to about 10\%.~\cite{btagging}
\end{itemize}

\section{Statistical Analysis}
\label{sec:stat}
\begin{figure}
\centering
\subfigure[]{
\resizebox{0.67\columnwidth}{!}{\includegraphics{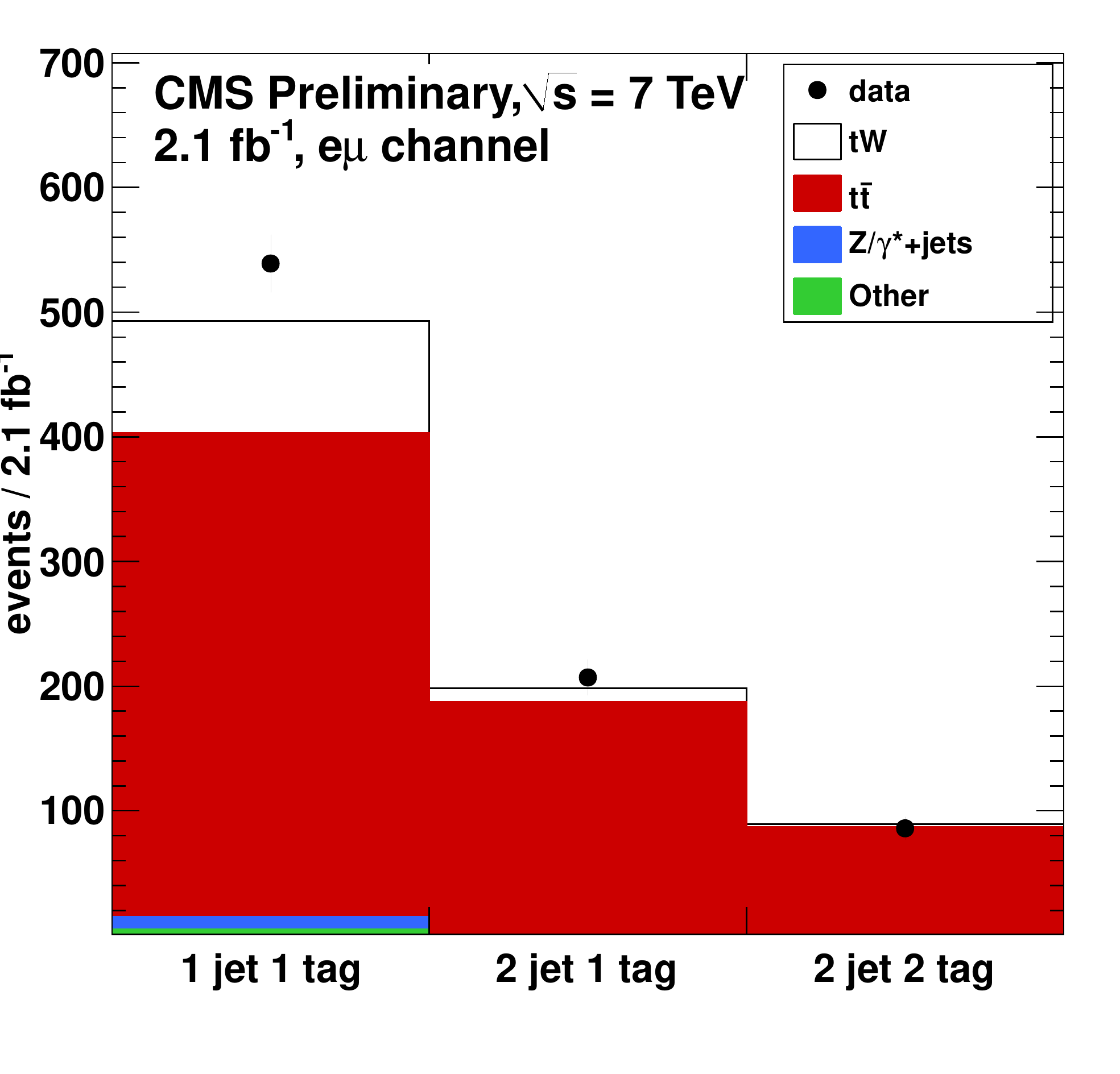}}
}
\subfigure[]{
\resizebox{0.67\columnwidth}{!}{\includegraphics{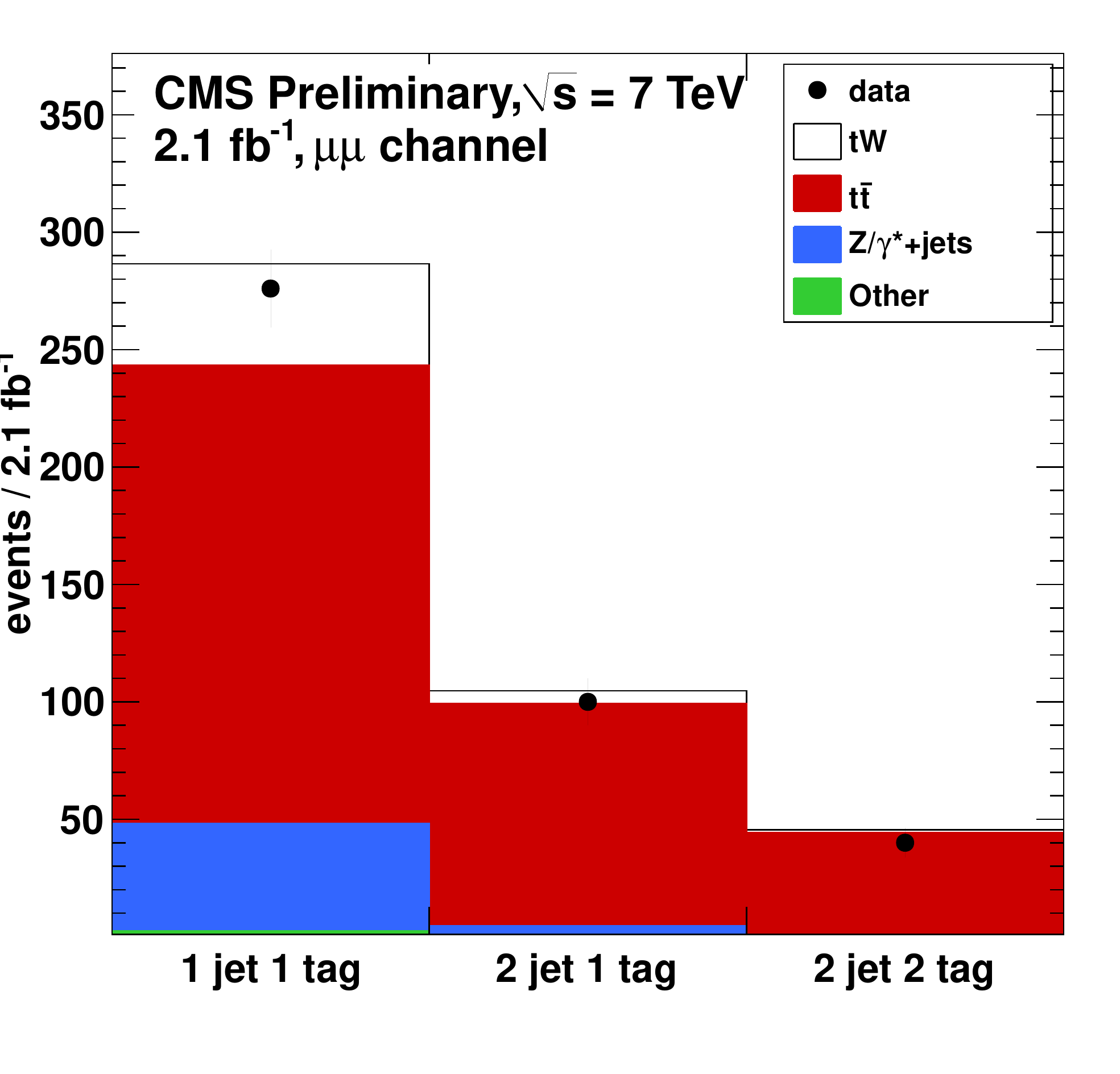}}
}
\subfigure[]{
\resizebox{0.67\columnwidth}{!}{\includegraphics{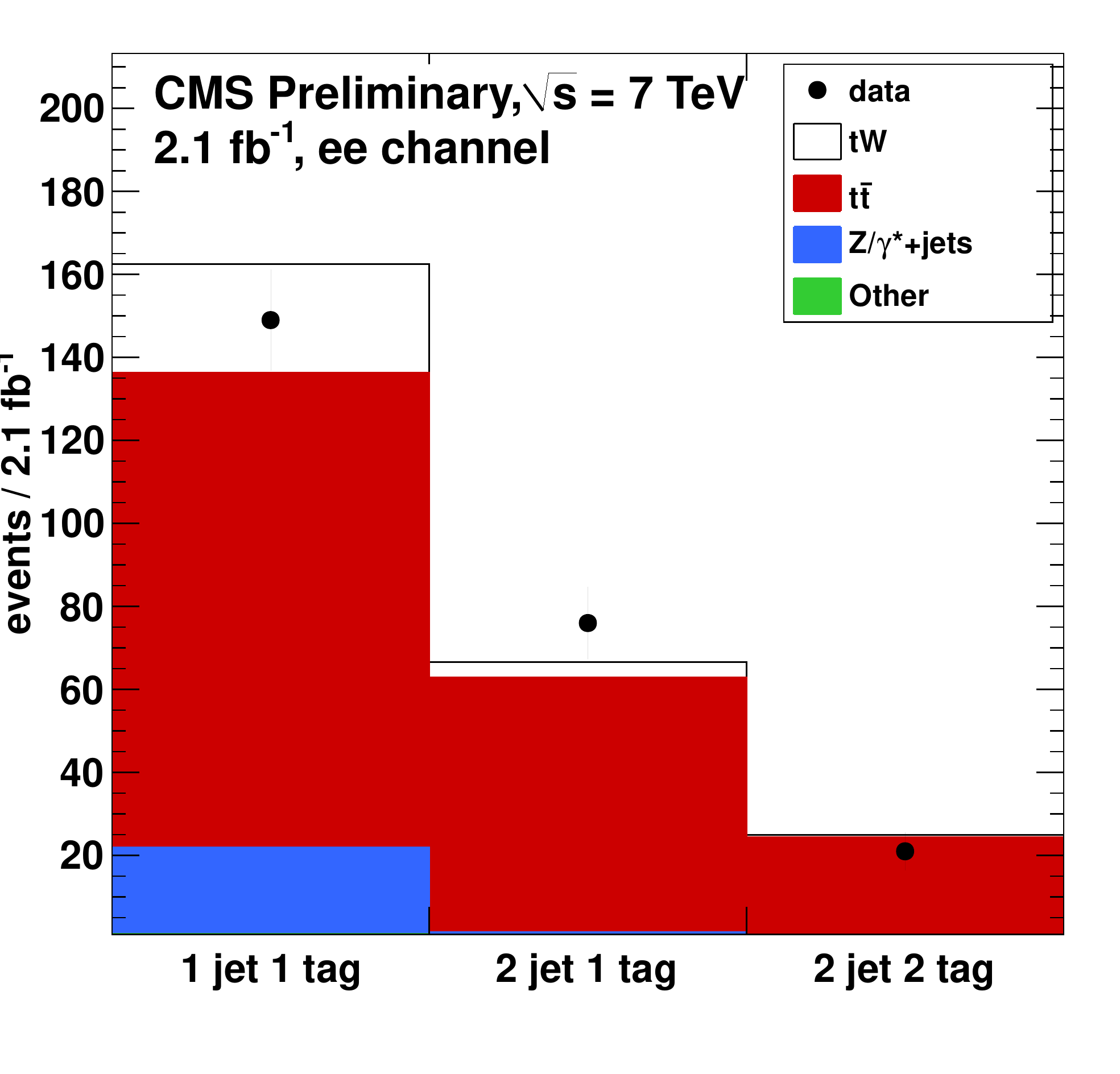}}
}
\subfigure[]{
\resizebox{0.67\columnwidth}{!}{\includegraphics{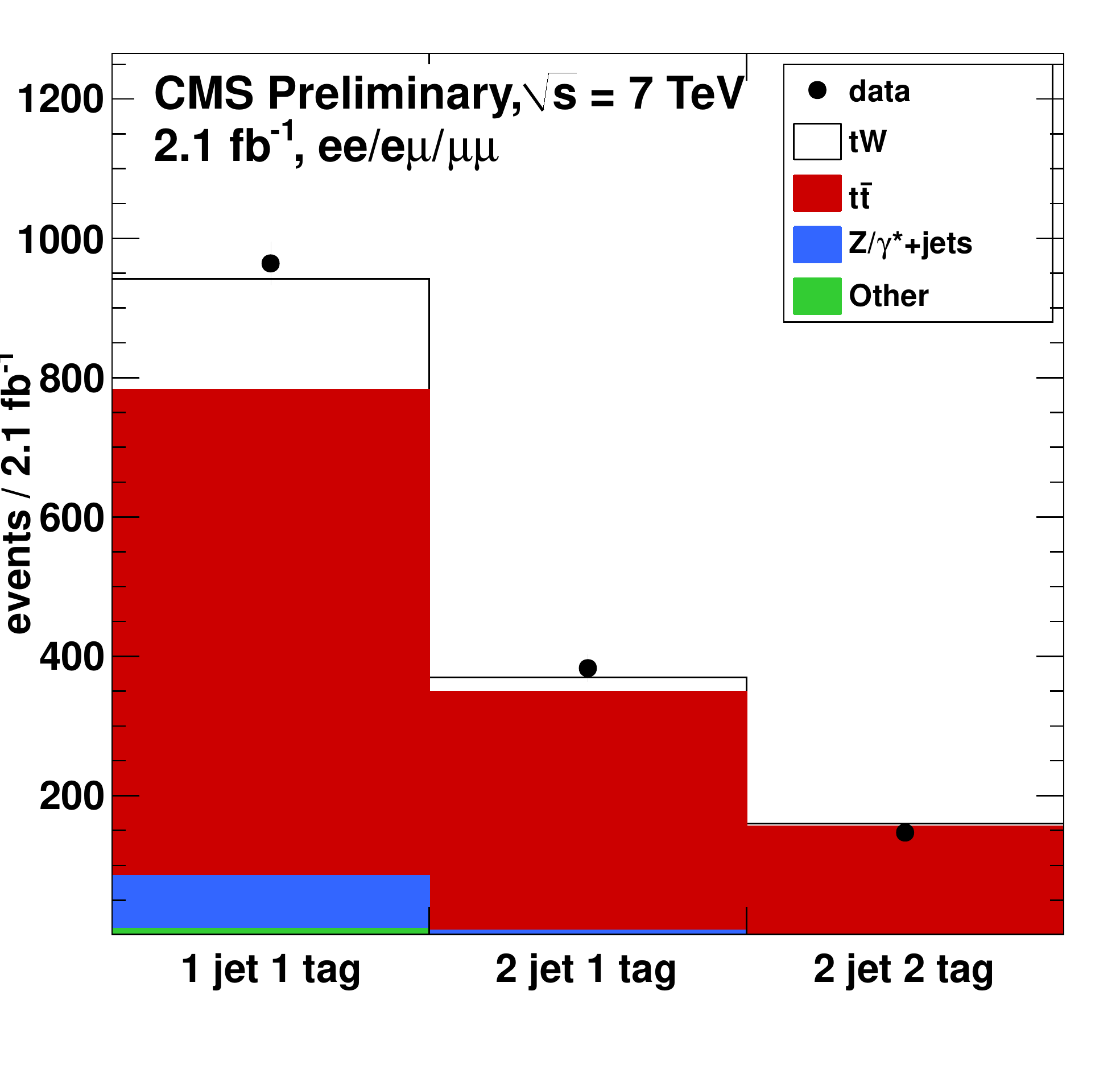}}
}
\caption{Oberved and predicted number of events in all lepton and jet / b-tagging categories.
The prediction is evaluated at the maximum of the likelihood function.}
\label{fig:9chan}
\end{figure}

To extract the cross section and significance, a Poisson counting model is used
in 9 channels: 3 lepton final states (ee/e$\mu$/$\mu\mu$), each with 3 jet / b-tag
multiplicities shown in Fig.\ \ref{fig:9chan}. The likelihood function is
the product of Poisson probabilities over all 9 channels:
\[
L(\vec p | \vec n) = \prod_{i=1}^9 \frac{\mu_i^{n_i} e^{\mu_i}}{n_i!}
\]
where $n_i$ is the number of observed events in channel $i$ and $\mu_i$ is the predicted event
yield which depends on the model parameters $\vec p$.
Systematic uncertainties are included as nuisance parameters as log-normal
uncertainties which change the predicted yield $\mu_i$ correlated across all channels.

The cross section is extracted using a profile likelihood technique.
The significance is evaluated using the
distribution of a profile likelihood ratio test statistic for toy experiments including no signal. For
toy generation, the nuisance parameters are drawn randomly according to their priors.

\section{Result}
The extracted single top tW cross section is
\[
\sigma_{\rm{tW}} = 22^{+9}_{-7} \rm{ pb}
\]
which is consistent with the Standard Model prediction.
The uncertainty includes the statistical and all systematic uncertainties discussed in Section~\ref{sec:syst}.

The probability that the observed excess of events is merely an upward fluctuation of background
with a vanishing tW cross section corresponds to $2.7\sigma$, the expected significance for
the Standard Model cross section is $1.8\pm 0.9\sigma$.

More details about this analysis can be found in~\cite{thispas}.

%

\end{document}